\documentstyle[twoside,fleqn,npb,epsfig]{article}
\newcommand{\AmS}{{\protect\the\textfont2
  A\kern-.1667em\lower.5ex\hbox{M}\kern-.125emS}}

\title{Hard exclusive production of two pions:
Dipion mass distributions in $\gamma^* N\to \pi\pi N$
and $\gamma^*\gamma\to \pi\pi$}

\author{M.V.\ Polyakov
\address{Petersburg Nuclear physics Institute, Theory Division,
        188350 Gatchina, Russia\\
and\\
Institut f\"ur Theoretische Physik II,
Ruhr--Universit\"at Bochum, D--44780 Bochum, Germany}
        \thanks{e-mail: maximp@tp2.ruhr-uni-bochum.de}}

\begin{document}

\begin{abstract}
The leading twist parametrization of dipion mass distribution in
hard exclusive reactions is proposed. Its parameters are related
to quark distributions (usual and skewed) in the pion and to
distributions amplitudes of mesons ($\pi$, $\rho$, etc.).  We show
that  measurements of the shape of dipion mass distribution in
hard exclusie reactions can give important information about
partonic structure of the pion. The expression for the amplitude
of the reaction $\gamma^*\gamma\to \pi\pi$ near the threshold in
terms of singlet quark distribution in the pion is presented.
\end{abstract}

% typeset front matter (including abstract)
\maketitle

\section{Dipion mass distribution in\\
 $\gamma^*+p\to \pi^+\pi^-+p $}

Recently it became possible to measure with good precision two pion
hard exclusive production in the deeply virtual photon fragmentation
region in the range of dipion masses $0.4\leq m_{\pi\pi}\leq 1.5$~GeV
(see talks at this conference \cite{exptalks} and
refs.~\cite{zeus,h1}).

Owing to QCD factorization theorem for hard exclusive reaction \cite{CFS}
the dependence of the amplitude  of the reactions

\begin{equation}
\gamma^*_L+T\to \pi\pi + T'
\label{proc2}
\end{equation}
on the dipion mass $m_{\pi\pi}$ factorizes at leading order
into the universal (independent of the target) factor:
\begin{equation}
\int_0^1 \frac{dz}{z}\,
\Phi^{I}(z,\zeta,m_{\pi\pi};  Q^2)\;.
\label{f1}
\end{equation}
Here $\Phi^{I}(z,\zeta,m_{\pi\pi};  Q^2)$ is two-pion light cone
distribution amplitude ($2\pi$DA), which depends on
$z$--longitudinal momentum carried by the quark, $\zeta$
characterizing the distribution of longitudinal momentum between
the two pions, and the invariant mass of produced pions
$m_{\pi\pi}$, superscript $I$ stands for isospin of produced pions
($I=0,1$). The dependence on the virtuality of the incident photon
$Q^2$ is governed by the evolution equation.  The $2\pi$DA's were
introduced recently in \cite{Ter} in the context of the QCD
description of the process $\gamma^\ast \gamma \to 2 \pi$, its
detailed properties were studied in \cite{MVP98}, we refer to
these papers for details.

For the process (\ref{proc2}) at small $x_{Bj}$
(see $e.g.$ recent measurements \cite{zeus,h1}) the production of two
pions in the isoscalar channel is strongly suppressed
relative to the isovector channel, because the former is
mediated by $C$-parity odd exchange.
 At asymptotically large $Q^2$ QCD
predicts the following simple form for the isovector $2\pi$DA
\cite{PW98,MVP98}:

\begin{equation}
\Phi^{I=1}_{as}(z,\zeta,m_{\pi\pi})=
6  z(1-z) (2\zeta-1)F_\pi(m_{\pi\pi})\; ,
\label{asym}
\end{equation}
where $F_\pi(m_{\pi\pi})$ is the pion e.m. form factor measured with high
precision in low energy experiments \cite{Barkov}. From
eqs.~(\ref{f1},\ref{asym}) we conclude that at asymptotically large
$Q^2$ QCD predicts unambiguously the shape of the dipion mass distribution:

\begin{equation}
\frac{dN(m_{\pi\pi})}{dm_{\pi\pi}^2}\propto
\biggl(1-\frac{4 m_\pi^2}{m_{\pi\pi}^2}\biggr)^{3/2}
 |F_{\pi}(m_{\pi\pi})|^2\, .
\label{asyshape}
\end{equation}

At non-asymptotic $Q^2$ the $2\pi$DA deviates from its asymptotic
form (\ref{asym}). This deviation can be described by a few parameters
which can be related to
quark distributions (skewed and usual) in the pion and to distribution
amplitudes of mesons ($\pi\rho$, etc. ), for details see \cite{MVP98}.
In this case we propose the following parametrization of dipion mass
distribution:
\begin{eqnarray}
\nonumber
\frac{dN(m_{\pi\pi})}{dm_{\pi\pi}^2}&\propto&
\biggl(1-\frac{4 m_\pi^2}{m_{\pi\pi}^2}\biggr)^{3/2}
|F_{\pi}(m_{\pi\pi})|^2 \\
&\times&\biggl(1+D_1(m_{\pi\pi},Q^2)\biggr)^2\\
\nonumber
&+&\frac 37 \
\biggl(1-\frac{4 m_\pi^2}{m_{\pi\pi}^2}\biggr)^{7/2}
|D_2(m_{\pi\pi},Q^2)|^2 ,
\label{shape}
\end{eqnarray}
where the functions
$D_{1,2}(m_{\pi\pi},Q^2)$  describe the deviation of the dipion
mass distribution from its asymptotic form (\ref{asyshape}) and
can be parametrized in the form:
\begin{eqnarray}
\nonumber
D_1(m_{\pi\pi},Q^2)&=&C_1(Q^2) e^{b_1m_{\pi\pi}^2}\\
&-&\frac{6 m_\pi^2}{m_{\pi\pi}^2}\
C_2(Q^2)
e^{b_2m_{\pi\pi}^2} \\
\nonumber
D_2(m_{\pi\pi},Q^2)&=&C_2(Q^2) e^{b_3 m_{\pi\pi}^2}\, .
\end{eqnarray}
The dependence of $C_{1,2}(Q^2)$ on $Q^2$ is governed by the
QCD evolution and in leading order is given by:

\begin{eqnarray}
C_{1,2}(Q^2)=C_{1,2}(\mu_0)
\biggl(
\frac{\alpha_s(Q^2)}{\alpha_s(\mu_0)}
\biggr)^{50/(99-6 n_f)}\, .
\end{eqnarray}
With increasing of $Q^2$ the parameters $C_{1,2}(Q^2)$ go
logarithmically to zero and one reproduces the asymptotic
formula (\ref{asyshape}).

The parameters $b_i$ are $Q^2$ independent but, in principle, $m_{\pi\pi}$
dependent. The latter dependence is fixed by $\pi\pi$ scattering
phase shifts, see \cite{MVP98}.

In derivation of eq.~(\ref{shape}) we neglect the production
of pions in $G$-waves and higher because it is suppressed by
powers of $1/\log(Q^2)$.

Interestingly that the parameters $C_{1,2}(\mu_0)$ and $b_i$ can be related
to important parameters of quark distributions in the pion and meson
distribution amplitudes \cite{MVP98}.
Using soft pion theorems for $2\pi$DA's \cite{MVP98} one can
express the second Gegenbauer moment of pion distribution amplitude
in terms of $C_{1,2}(\mu_0)$:
\begin{equation}
a_2^{(\pi)}(\mu_0)=C_{1}(\mu_0)+ C_{2}(\mu_0)
\;.
\end{equation}
If one additionally uses the dispersion relation derived in \cite{MVP98}
one gets the expression for the second Gegenbauer moment of $\rho$-meson
distribution amplitude:
\begin{equation}
a_2^{(\rho)}(\mu_0)\approx
C_{1}(\mu_0) \exp(b_1 m_\rho^2)
\;.
\end{equation}
The crossing relations \cite{MVP98} (see also \cite{dub,PW99})
allow to relate the parameter
$C_{2}(\mu_0)$ to the third Mellin moment of quark distribution
in the pion at normalization point $\mu_0$

\begin{equation}
\int_{0}^1 dx\ x^{2} (u^{\pi^+}(x)-\bar u^{\pi^+}(x))=
\frac{6}{7}\  C_{2}(\mu_0)\, .
\end{equation}

In analysis of experiments on two pion diffractive production
off nucleon (see e.g. \cite{zeus,h1}) the dipion mass distribution
is usually fitted  by, for example, S\"oding parametrization, which
takes into account rescattering of produced pions on final
nucleon. Let us note however that in the case of hard ($Q^2\to \infty$)
diffractive production the final state interaction of pions with
residual nucleon is suppressed by powers of $1/Q^2$ relative to the
leading twist amplitude. Also the parameters of S\"oding
(or Ross-Stodolsky) parametrization are not related to any
fundamental parameter of QCD.
Here we proposed alternative leading-twist parametrization
(\ref{shape}) describing the so-called ``skewing" of two pion
spectrum. The advantage of our parametrization is that its parameters can
be related to fundamental quantities in QCD: quark distributions
in the pion and meson distribution amplitudes.

\section{Reaction $\gamma^* \gamma\to\pi\pi$
 near threshold}

Another example of hard exclusive reaction where the $2\pi$DA
enters is the reaction $\gamma^* \gamma\to\pi\pi$ with
$Q^2$ of the photon large and the invariant mass of produced
pions is small. In refs.~\cite{Ter,Freund} it was demostrated that
this process is amenable to QCD description. The amplitude
in the leading order can be expressed in terms of isoscalar
$2\pi$DA \cite{Ter}:
\begin{eqnarray}
M_{\mu\nu}=\sum_f e_f^2\  g^\perp_{\mu\nu}
\int_0^1 \frac{dz}{z}\,
\Phi^{I=0}(z,\zeta,m_{\pi\pi};  Q^2)\;.
\label{amptp}
\end{eqnarray}
Using the crossing relations and soft pion theorems we
derive the expression for the amplitude $M_{\mu\nu}$ close to the
threshold $m_{\pi\pi}=2m_\pi$ in terms of singlet quark
distribution in the pion. To this end we make one additional
approximation: that production of two pions in the state with angular
momentum $l$ is dominated by the operators with conformal spin
$l-1$ and $l+1$.
The higher $Q^2$ the more this approximation is
justified, see for details \cite{else}.
The final expression for the integral over $z$ entering eq.~(\ref{amptp})
has the form:
\begin{eqnarray}
\nonumber
&&\int_0^1 \frac{dz}{z}\,
\Phi^{I=0}(z,\zeta,m_{\pi\pi}\approx 2 m_\pi)=\\
\nonumber
&&\int_0^1 \frac{dx}{x}Q(x)\, \biggl(
\frac{1-x^2}{\sqrt{(1+x)^2-4\zeta x}}\\
&&+
\frac{1-x^2}{\sqrt{(1-x)^2+4\zeta x}}-2
\biggl)\, ,
\label{dual}
\end{eqnarray}
where the function $Q(x)$ is related to the singlet quark distribution
in the pion:
\begin{eqnarray}
\nonumber
Q(x)=2 q_\pi(x)-x\int_x^1 \frac{dy}{y^2}q_\pi(y)\, ,
\end{eqnarray}
with
\begin{eqnarray}
\nonumber
 q_\pi(x)=\frac{1}{N_f} \sum_f (q_f(x)+\bar q_f(x))
\end{eqnarray}
Note that the normalization points of $ q_\pi(x)$ and
$\Phi^{I=0}(z,\zeta,m_{\pi\pi})$ are the same.

Using the eq.~(\ref{dual}) and dispersion relations
derived in ref.~\cite{MVP98} we can write the following
expression for the amplitude (\ref{amptp}) keeping only $S-$
and $D-$ waves:

\begin{eqnarray}
\nonumber
\lefteqn{
M_{\mu\nu} = \frac{5}{9} e^2  g^\perp_{\mu\nu}
\biggl[
A_0\ f_0(m_{\pi\pi}) e^{i\delta_0^0(m_{\pi\pi})}\ P_0(\cos\theta_{\rm cm})+
} && \\
&& A_2\
f_2(m_{\pi\pi}) e^{i\delta_2^0(m_{\pi\pi})}\ P_2(\cos\theta_{\rm cm})
\biggr]\, ,
\label{sd}
\end{eqnarray}
where the  $A_{0}$ and $A_{2}$ are given in terms of
singlet quark distribution in the pion. For $A_0$ we give the full
expression:

\begin{eqnarray}
&&A_0=-\frac{1}{\beta} \int_0^1 dx Q(x)\, \biggl(
\frac{2\beta}{x}+\\
\nonumber
&&\frac{1-x^2}{x^2}(
\sqrt{1-2\beta x+x^2}-\sqrt{1+2\beta x+x^2}
)\biggr) ,
\label{a0}
\end{eqnarray}
where $\beta$ is the velocity of produced pion in centre of mass
frame
\begin{eqnarray}
\nonumber
\beta=\sqrt{1-\frac{4 m_\pi^2}{m_{\pi\pi}^2}}\, .
\end{eqnarray}
The scattering angle in c.m. frame is:

\begin{eqnarray}
\nonumber
\cos \theta_{\rm cm}=\frac{2\zeta-1}{\beta}\,.
\end{eqnarray}

In two limiting cases: nonrelativistic pions ($\beta\to 0$) and
ultrarelativistic ($\beta\to 1$) that we have for $A_0$:

\begin{eqnarray}
\nonumber
A_0 &=&\int_0^1 dx Q(x)
\left\{ \begin{array}{ll}  \frac{2}{x}(\frac{1-x^2}{\sqrt{1+x^2}}-1)
& \beta \rightarrow 0 \\  -2x
& \beta \rightarrow 1 \end{array} \right.
\end{eqnarray}
The full expression for $A_2$ is more complicated and we give
here only its limiting cases:
\begin{eqnarray}
\nonumber
A_2 &=&\int_0^1 dx Q(x)
\left\{ \begin{array}{ll}  \frac{2x(1-x^2)}{(1+x^2)^{5/2}}
\beta^2
& \beta \rightarrow 0 \\  2x(1-x^2)
& \beta \rightarrow 1 \end{array} \right.
\end{eqnarray}

The functions $f_{0,2}(m_{\pi\pi})$ in eq.~(\ref{sd}) can be
related to $\pi\pi$ phase shifts $\delta_0^0(m_{\pi\pi})$ and
$\delta_2^0(m_{\pi\pi})$ using Watson theorem and dispersion
relations derived in \cite{MVP98}:

\begin{eqnarray}
&&\log f_l(m_{\pi\pi})=\\
\nonumber
&&m_{\pi\pi}^2 \biggl(
b_l+
\frac{m_{\pi\pi}^{2}}{\pi}
{\rm Re} \int_{4m_\pi^2}^\infty
ds \frac{\delta_l^0(s)}{s^2(s-m_{\pi\pi}^2-i0)}
\biggr),
\label{omnes}
\end{eqnarray}
where the constants $b_{l}$ can be estimated using low-energy
models of QCD. For example, the estimate in the instanton model of the QCD
vacuum gives \cite{MVP98}
$$
b_0=b_2=\frac{N_c}{48\pi^2 f_\pi^2}\approx 0.73 \ {\rm GeV}^{-2} \, .
$$

Let us note that the expression (\ref{omnes}), strictly speaking,
is valid only in the elastic region
($4 m_\pi^2 \leq m_{\pi\pi}^2 \leq 16 m_\pi^2$).
It is rather easy to extend its range of applicability including
the contributions of higher intermediate states  (probably the most
important is the contribution of $K\bar K$) in the dispersion relations.

\section{Conclusions}
We showed on two examples of hard exclusive reactions that
the measurements of the shape of dipion mass distributions
in these reactions can provide us with important information
on partonic structure of the pion and two-pion resonances.

\end{document}